&pdflatex
\documentclass[12pt]{article}
\usepackage{amsmath}
\usepackage{graphicx,psfrag,epsf}
\usepackage{enumerate}
\usepackage{natbib}
\usepackage{url} 
\usepackage{color}

\newcommand{\blind}{0}

\begin{document}

\bibliographystyle{natbib}

\def\spacingset#1{\renewcommand{\baselinestretch}%
{#1}\small\normalsize} \spacingset{1}


\if0\blind
{
  \title{\bf Abandon Statistical Significance}
  \author{{\em Forthcoming in } The American Statistician \\
  \phantom{a} \\
  Blakeley B. McShane\thanks{Correspondence concerning this article should be addressed to Blakeley B. McShane, Marketing Department, Kellogg School of Management, Northwestern University, 2211 Campus Drive, Evanston, IL 60208. E-mail: b-mcshane@kellogg.northwestern.edu. We thank the National Science Foundation, the Institute for Education Sciences, and the Office of Naval Research for partial support of Andrew Gelman's work.}\hspace{.2cm}, Northwestern University \\
    David Gal, University of Illinois at Chicago \\
    Andrew Gelman, Columbia University \\
    Christian Robert, Universit\'e Paris-Dauphine \\
    Jennifer L. Tackett, Northwestern University}
  \maketitle
} \fi

\if1\blind
{
  \bigskip
  \bigskip
  \bigskip
  \begin{center}
    {\LARGE\bf Abandon Statistical Significance}
\end{center}
  \medskip
} \fi

\bigskip
\vspace{-0.5in}
\begin{abstract}
We discuss problems the null hypothesis significance testing (NHST) paradigm poses for replication and more broadly in the biomedical and social sciences as well as how these problems remain unresolved by proposals involving modified $p$-value thresholds, confidence intervals, and Bayes factors. We then discuss our own proposal, which is to abandon statistical significance. We recommend dropping the NHST paradigm---and the $p$-value thresholds intrinsic to it---as the default statistical paradigm for research, publication, and discovery in the biomedical and social sciences. Specifically, we propose that the $p$-value be demoted from its threshold screening role and instead, treated continuously, be considered along with currently subordinate factors (e.g., related prior evidence, plausibility of mechanism, study design and data quality, real world costs and benefits, novelty of finding, and other factors that vary by research domain) as just one among many pieces of evidence. We have no desire to ``ban'' $p$-values or other purely statistical measures. Rather, we believe that such measures should not be thresholded and that, thresholded or not, they should not take priority over the currently subordinate factors. We also argue that it seldom makes sense to calibrate evidence as a function of $p$-values or other purely statistical measures. We offer recommendations for how our proposal can be implemented in the scientific publication process as well as in statistical decision making more broadly.
\end{abstract}

\noindent%
\vfill

\newpage
\spacingset{1.45} 

\section{The Status Quo and Two Alternatives}

The biomedical and social sciences are facing a widespread crisis, with published findings failing to replicate at an alarming rate. Often, such failures to replicate are associated with claims of huge effects from subtle, sometimes even preposterous, interventions. Further, the primary evidence adduced for these claims is one or more comparisons that are anointed ``statistically significant''---typically defined as comparisons with $p$-values less than the conventional 0.05 threshold relative to the sharp point null hypothesis of zero effect and zero systematic error. 

Indeed, the {\em status quo} is that $p < 0.05$ is deemed as strong evidence in favor of a scientific theory and is required not only for a result to be published but even for it to be taken seriously. Specifically, statistical significance serves as a lexicographic decision rule whereby any result is first required to have a $p$-value that attains the 0.05 threshold and only then is consideration---often scant---given to such factors as related prior evidence, plausibility of mechanism, study design and data quality, real world costs and benefits, novelty of finding, and other factors that vary by research domain (for want of a better term, we hereafter refer to these collectively as the {\em currently subordinate factors}).

Traditionally, the $p < 0.05$ rule has been considered a safeguard against noise-chasing and thus a guarantor of replicability. However, in recent years, a series of well-publicized examples (e.g., \cite{CarCudYap10} and \cite{Bem11}) coupled with theoretical work has made it clear that statistical significance can easily be obtained from pure noise. Consequently, low replication rates are to be expected given existing scientific practices \cite[]{Ioa05,SmaMcE16}, and calls for reform, which are not new (see, for example, \cite{Meehl78}), have become insistent.

One proposal, suggested by Daniel Benjamin and seventy-one coauthors including distinguished scholars from a wide variety of fields, is to {\em redefine} statistical significance, ``to change the default $p$-value threshold for statistical significance for claims of new discoveries from 0.05 to 0.005'' \cite[]{Benjetal18}. While, as they note, ``changing the $p$-value threshold is simple, aligns with the training undertaken by many researchers, and might quickly achieve broad acceptance,'' we believe this ``quick fix,'' this ``dam to contain the flood'' in the words of a prominent member of the seventy-two \cite[]{Resn17}, is insufficient to overcome current difficulties with replication. Instead, we believe it opportune to proceed immediately with other measures, perhaps more radical and more difficult but also more principled and more permanent.

In particular, we propose to {\em abandon} statistical significance, to drop the null hypothesis significance testing (NHST) paradigm---and the $p$-value thresholds intrinsic to it---as the default statistical paradigm for research, publication, and discovery in the biomedical and social sciences. Specifically, rather than allowing statistical significance as determined by $p < 0.05$ (or some other threshold whether based on $p$-values, confidence intervals, Bayes factors, or some other purely statistical measure) to serve as a lexicographic decision rule in scientific publication and statistical decision making more broadly, we propose that the $p$-value be demoted from its threshold screening role and instead, treated continuously, be considered along with the currently subordinate factors as just one among many pieces of evidence.

To be clear, we have no desire to ``ban'' $p$-values or other purely statistical measures. Rather, we believe that such measures should not be thresholded and that, thresholded or not, they should not take priority over the currently subordinate factors. We also argue that it seldom makes sense to calibrate evidence as a function of $p$-values or other purely statistical measures.

In the remainder of this paper, we discuss general problems with NHST that motivate our proposal to abandon statistical significance and that remain unresolved by the \cite{Benjetal18} proposal. We then discuss problems specific to the \cite{Benjetal18} proposal. We next offer recommendations for how, in practice, the $p$-value can be demoted from its threshold screening role and instead be considered as just one among many pieces of evidence in the scientific publication process as well as in statistical decision making more broadly. We conclude with a brief discussion.

\section{Problems General to Null Hypothesis Significance Testing}

\subsection{Preface}

As noted, the NHST paradigm upon which the status quo and the \cite{Benjetal18} proposal rest is the default statistical paradigm for research, publication, and discovery in the biomedical and social sciences (see, for example, \cite{MorHen70,Gig87,SawPet83,McCZil96,Gill99,AndBurTho00,Gig04,Hubbard04}). Despite this, it has been roundly criticized both inside and outside of statistics over the decades (see, for example, \cite{Ros60,Bakan66,Meehl78,SerLap93,Coh94,McCZil96,Schm96,Hun97,Gill99,Gig04,GigKraVit04,Briggs16,McSGal16,McSGal17}). Indeed, the breadth of literature on this topic across time and fields makes a complete review intractable. Consequently, we focus on what we view as among the most important criticisms of NHST for the biomedical and social sciences.

\subsection{Implausible Null Hypothesis}

In the biomedical and social sciences, effects are typically small and vary considerably across people and contexts. In addition, measurements are often highly variable and only indirectly related to underlying constructs of interest; thus, even when sample sizes are large, the possibilities of systematic bias and variation can result in the equivalent of small or unrepresentative samples. Consequently, estimates from any single study are themselves generally noisy. Nonetheless, the single study is typically the fundamental unit of analysis.

In addition, the null hypothesis employed in the overwhelming majority of applications is the sharp point null hypothesis of zero effect---that is, no difference among two or more treatments or groups---and zero systematic error---which encompasses both the adequacy of the statistical model used to compute the $p$-value (e.g., in terms of functional form and distributional assumptions) as well as any and all forms of systematic or non-sampling error which vary by field but include measurement error; problems with reliability and validity; biased samples; non-random treatment assignment; missingness; non-response; failure of double-blinding; non-compliance; and confounding.

The combination of these features of the biomedical and social sciences and this sharp point null hypothesis of zero effect and zero systematic error is highly problematic. Specifically, because effects are generally small and variable, the assumption of zero effect is false. Further, even were the assumption of zero effect true for some phenomenon, the effect under consideration in any study designed to examine this phenomenon would not be zero because measurements are generally noisy and systematically so. Consequently, the sharp point null hypothesis of zero effect and zero systematic error employed in the overwhelming majority of applications is implausible \cite[]{Berk38,EdwLinSav63,Bakan66,Meehl90,Tukey91,Coh94,GelCarSte14,McSBoc14,Gel15} and thus uninteresting. 

These problems are exacerbated under a lexicographic decision rule for publication as per the status quo and the \cite{Benjetal18} proposal. Specifically, because noisy estimates that attain statistical significance are upwardly biased in magnitude (potentially to a large degree) and often of the wrong sign \cite[]{GelCar14}, a lexicographic decision rule results in a tarnished literature. In addition, because many smaller, less resource-intensive, noisier studies are more likely to yield (or can be made more likely to yield; \cite{SimNelSim11}) one or more statistically significant results than fewer larger, more resource-intensive, better studies, a lexicographic decision rule at least indirectly encourages the former over the latter. These issues are compounded when researchers engage in multiple comparisons---whether actual or potential (i.e., the ``garden of forking paths''; \cite{GelLok14}).

In sum, various features of the biomedical and social sciences---for example, small and variable effects, systematic error, noisy measurements, a lexicographic decision rule for publication, and research practices---make NHST and in particular the sharp point null hypothesis of zero effect and zero systematic error particularly poorly suited for these domains.

\subsection{Categorization of Evidence}

NHST is associated with a number of problems related to the dichotomization of evidence into the different categories ``statistically significant'' and ``not statistically significant'' (or, sometimes, trichotomization with ``marginally significant'' as an intermediate category) depending upon where the $p$-value stands relative to certain conventional thresholds. Indeed, one well-known criticism of the NHST paradigm is that the conventional 0.05 threshold---or for that matter any other one---is entirely arbitrary \cite[]{Fish26,YulKen50,Cram55,Coch76,CowDav89}. 

A related line of criticism suggests that the problem is with having a threshold in the first place: the dichotomization (or trichotomization) of evidence into different categories of statistical significance itself has ``no ontological basis'' \cite[]{RosRos89}. Specifically, \cite{RosRos89} note that ``from an ontological viewpoint, there is no sharp line between a `significant' and a `nonsignificant' difference; significance in statistics...varies continuously between extremes'' and thus advocate ``view[ing] the strength of evidence for or against the null as a fairly continuous function of the magnitude of $p$.''

While we agree treating the $p$-value continuously rather than in a thresholded manner constitutes an improvement, we  go further and argue that it seldom makes sense to calibrate evidence as a function of the $p$-value. We hold this for at least three reasons. First, and in our view the most important, it seldom makes sense because the $p$-value is, in the overwhelming majority of applications, defined relative to the generally implausible and uninteresting sharp point null hypothesis of zero effect and zero systematic error. Second, because it is a poor measure of the evidence for or against a statistical hypothesis \cite[]{EdwLinSav63,BerSel87,Coh94,HubLin08}. Third, because it tests the hypothesis that one or more model parameters equal the tested values---but only given all other model assumptions. These other assumptions---in particular, zero systematic error---seldom hold (or are at least far from given) in the biomedical and social sciences. Consequently, ``a small $p$-value only signals that there may be a problem with at least one assumption, without saying which one. Asymmetrically, a large $p$-value only means that this particular test did not detect a problem---perhaps because there is none, or because the test is insensitive to the problems, or because biases and random errors largely canceled each other out'' \cite[]{Gree17}. We note similar considerations hold for other purely statistical measures.

\subsection{Erroneous Scientific Reasoning}

The NHST paradigm and the $p$-value thresholds intrinsic to it are not only problematic in and of themselves but also they routinely result in erroneous scientific reasoning. For example, researchers typically take the rejection of the sharp point null hypothesis of zero effect and zero systematic error as positive or even definitive evidence in favor of some preferred alternative hypothesis---a logical fallacy. In addition, they often make scientific conclusions largely if not entirely based on whether or not a $p$-value crosses the 0.05 threshold instead of taking a more holistic view of the evidence that includes the consideration of the currently subordinate factors. Further, they often confuse statistical significance and practical importance (see, for example, \cite{Free93}). Finally, they often incorrectly believe a result with a $p$-value below 0.05 is evidence that a relationship is causal \cite[]{Holmetal01}.

In addition, because the assignment of evidence to different categories (e.g., statistically significant and not statistically significant) is a strong inducement to the conclusion that the items thusly assigned are categorically different, NHST encourages researchers to engage in dichotomous thinking, that is, to interpret evidence dichotomously rather than continuously. Specifically, researchers interpret evidence that reaches the conventional threshold for statistical significance as a demonstration of a difference, and, in contrast, they interpret evidence that fails to reach this threshold as a demonstration of no difference.

An example of erroneous reasoning resulting from dichotomous thinking is provided by \cite{GelSte06} who show that applied researchers often fail to appreciate that ``the difference between `significant' and `not significant' is not itself statistically significant.'' Additional examples are provided by \cite{McSGal16} who show that researchers across a wide variety of fields including medicine, epidemiology, cognitive science, psychology, and economics (i) interpret $p$-values dichotomously rather than continuously, focusing solely on whether or not the $p$-value is below 0.05 rather than the magnitude of the $p$-value; (ii) fixate on $p$-values even when they are irrelevant, for example when asked about descriptive statistics; and (iii) ignore other evidence, for example the magnitude of treatment differences. \cite{McSGal17} show that even statisticians are susceptible to these errors.

\subsection{Misinterpretation of the $p$-value}

A final criticism against the NHST paradigm pertains to common misinterpretations of the $p$-value. While formally defined as the probability of observing data as extreme or more extreme than that actually observed assuming the null hypothesis is true, the $p$-value has often been misinterpreted as, {\em inter alia}, (i) the probability that the null hypothesis is true, (ii) one minus the probability that the alternative hypothesis is true, or (iii) one minus the probability of replication. For example, \cite{Gig04} reports an example of research conducted on psychology professors, lecturers, teaching assistants, and students. Subjects were given the result of a simple $t$-test of two independent means ($t=2.7$, $df=18$, $p=0.01$) and were asked six true or false questions based on the result and designed to test common misinterpretations of the $p$-value. All six of the statements were false and, despite the fact that the study materials noted ``several or none of the statements may be correct,'' (i) none of the forty-four students, (ii) only four of the thirty-nine professors and lectures who did not teach statistics, and (iii) only six of the thirty professors and lectures who did teach statistics marked all as false (members of each group marked an average of 3.5, 4.0, and 4.1 statements respectively as false). For related results, see \cite{Oakes86}, \cite{Coh94}, \cite{HalKra02}, and \cite{Gig18}.

\section{Problems Specific to the \cite{Benjetal18} Proposal}

Beyond concerns about the NHST paradigm upon which the status quo and the \cite{Benjetal18} proposal rest, there are additional problems specific to the latter proposal. First, \cite{Benjetal18} propose the 0.005 threshold because it (i) ``corresponds to Bayes factors between approximately 14 and 26'' in favor of the alternative hypothesis and (ii) ``would reduce the false positive rate to levels we judge to be reasonable.'' However, little to no justification is provided for either of these choices of levels.

Second, \cite{Benjetal18} ``restrict [their] recommendation to claims of discovery of new effects'' which is problematic for at least two reasons. First, the proposed policy is rendered entirely impractical because they fail to define what constitutes a new effect; this is especially so in domains where research is believed to be incremental and cumulative. Second, the proposed policy would lead to incoherence when applied to replication---the very issue their proposal is meant to address. In particular, the order in which two independent studies of a common phenomenon are conducted ought to be irrelevant but is not under the \cite{Benjetal18} proposal. Specifically, given one study with $p < 0.005$ and another with $p \in (0.005, 0.05)$, it would matter crucially which study was conducted first (and thus was ``new'') under the definition of replication employed in practice (i.e., a subsequent study is considered to successfully replicate a prior study if either both fail to attain statistical significance or both attain statistical significance and are directionally consistent): the second (replication) study would be deemed a success under the \cite{Benjetal18} proposal if the first study was the $p < 0.005$ study but a failure otherwise.

Third, the fact that uncorrected multiple comparisons---both actual and potential---are the norm in applied research strictly speaking invalidates all $p$-values outside those from studies with preregistered protocols and data analysis procedures. This concern is acknowledged by \cite{Benjetal18}. Nonetheless, what goes unacknowledged is that even with preregistration, $p$-values can be invalidated if the underlying model that generated the $p$-value is misspecified in an important manner.

Fourth, the mathematical justification underlying the \cite{Benjetal18} proposal has come under no small amount of criticism. Specifically, the uniformly most powerful Bayesian tests (UMPBTs) that underlie the proposal were introduced and defended by \cite{John13} in parallel with his call in \cite{John13b}---and now repeated in \cite{Benjetal18}---to use 0.005 as the new threshold. We see a number of concerns with UMPBTs that we discuss in Appendix \ref{umpbt}. Perhaps most relevant for the biomedical and social sciences, the UMPBT approach is deeply entrenched in the century-old Neyman-Pearson formalism of binary decisions and 0-1 loss functions which does not in general map, even in an approximate way, to processes of scientific learning or costs and benefits. Consequently, the logic underlying the proposal to move to a lower $p$-value threshold avoids firmly confronting the nature of the issue: any such rule implicitly expresses a particular tradeoff between Type I and Type II error, but in reality this tradeoff should depend on the costs, benefits, and probabilities of all outcomes \cite[]{GelRob14} which in turn depend on the problem at hand and which vary tremendously across studies and domains.

More speculatively, we are not convinced the more stringent 0.005 threshold for statistical significance would be helpful. In the short term, it could reduce the flow of low quality work that is currently polluting even top journals. In the medium term, it could motivate researchers to perform higher-quality work that is more likely to crack the 0.005 barrier. On the other hand, it could lead to even more overconfidence in results that do get published as well as a concomitant greater exaggeration of the effect sizes associated with such results. It could also lead to the discounting of important findings that happen not to reach the more stringent threshold.  In sum, we have no idea whether implementation of the proposed 0.005 threshold would improve or degrade the state of science as we can envision both positive and negative outcomes resulting from it. Ultimately, while this question may be interesting if difficult to answer, we view it as outside our purview because we believe that thresholds whether based on $p$-values or other purely statistical measures are a bad idea.

Perhaps curiously, we do not necessarily expect that \cite{Benjetal18} would disagree with our criticism that their proposal is insufficient to overcome current difficulties with replication (or perhaps even with our own proposal to abandon statistical significance). After all, they ``restrict [their] recommendation to claims of discovery of new effects''  and recognize that ``the choice of any particular threshold is arbitrary'' and ``should depend on the prior odds that the null hypothesis is true, the number of hypotheses tested, the study design, the relative cost of Type I versus Type II errors, and other factors that vary by research topic.'' Indeed, ``many of [the authors] agree that there are better approaches to statistical analyses than null hypothesis significance testing.''

\section{Abandoning Statistical Significance}

\subsection{Summation and Recommendations}

What can be done? Statistics is hard, especially when effects are small and variable and measurements are noisy. There are no quick fixes. Proposals such as changing the default $p$-value threshold for statistical significance, employing confidence intervals with a focus on whether or not they contain zero, or employing Bayes factors along with conventional classifications for evaluating the strength of evidence suffer from the same or similar issues as the current use of $p$-values with the 0.05 threshold. In particular, each implicitly or explicitly categorizes evidence based on thresholds relative to the generally implausible and uninteresting sharp point null hypothesis of zero effect and zero systematic error. Further, each is a purely statistical measure that fails to take a more holistic view of the evidence that includes the consideration of the currently subordinate factors, that is, related prior evidence, plausibility of mechanism, study design and data quality, real world costs and benefits, novelty of finding, and other factors that vary by research domain.

In brief, each is a form of statistical alchemy that falsely promises to transmute randomness into certainty, an ``uncertainty laundering'' \cite[]{Gel16} that begins with data and concludes with dichotomous declarations of truth or falsity---binary statements about there being ``an effect'' or ``no effect''---based on some $p$-value or other statistical threshold being attained. A critical first step forward is to begin accepting uncertainty and embracing variation in effects \cite[]{Car16,Gel16} and recognizing that we can learn much (indeed, more) about the world by forsaking the false promise of certainty offered by such dichotomization.

Towards this end, we offer recommendations for how, in practice, the $p$-value be demoted from its threshold screening role and instead, treated continuously, be considered along with the currently subordinate factors as just one among many pieces of evidence. First, we recommend authors use the currently subordinate factors to motivate their data collection, statistical analysis, interpretation of results, writing, and related matters; we also recommend they analyze and report all of their data and relevant results. Second, we recommend editors and reviewers explicitly evaluate papers with regard to not only purely statistical measures but also the currently subordinate factors.

As a highly inter-disciplinary research team with representation from statistics, political science, psychology, and consumer behavior, we are acutely aware that the implementation of our broad recommendations will and ought to vary tremendously across---and even within---domains. Further, we are not so supercilious to believe that we, by ourselves, are capable of providing concrete and specific guidance on the application of these recommendations across all or perhaps even any of these domains. Indeed, we do not believe a ``template'' for our recommendations is possible or desirable. In fact, such a template could even be dangerous in that it might result in a rote and recipe-like application of our recommendations that would not be entirely dissimilar to, even if perhaps less harmful than, the current practice of rote and recipe-like application of NHST. To those who might argue that, without such a template, our recommendations are unrealistic or unlikely to be adopted in practice, we reiterate that statistics is hard and a formulaic approach to statistics is a principal cause of the current replication crisis. It is for these reasons we advocate this more radical and more difficult but also more principled and more permanent approach. Nonetheless, we below suggest some broad principles that show how our recommendations might be applied. We also provide a case study in Appendix \ref{case}.

\subsection{For Authors}

We recommend authors use the currently subordinate factors to motivate their data collection, statistical analysis, interpretation of results, writing, and related matters; we also recommend they analyze and report all of their data and relevant results rather than focusing on single comparisons that attain some $p$-value or other statistical threshold. 

One specific operationalization of the first part of our recommendation might be to include in their manuscripts a section that directly addresses how each of the currently subordinate factors motivated their various decisions regarding data collection, statistical analysis, interpretation of results, and writing in the context of the totality of the data and results. Such a section could, for example, discuss study design in the context of subject-matter knowledge and expectations of effect sizes as discussed by \cite{GelCar14}. It could also discuss the plausibility of the mechanism by (i) formalizing the hypothesized mechanism for the effect in question and expounding on the various components of it, (ii) clarifying which components were measured and analyzed in the study, and (iii) discussing aspects of the results that support as well as those that undermine the hypothesized mechanism.

One might think that that the second part of our recommendation---to analyze and report all of the data and relevant results---is such a fundamental principle of science that it need hardly be mentioned. However, this is not the case! As discussed above, the status quo in scientific publication is a lexicographic decision rule whereby $p < 0.05$ is virtually always required for a result to be published and, while there are some exceptions, standard practice is to focus on such results and to not report all relevant findings. 

Given the current state of practice, authors may not have a sense for how they might go about this. Rather than attempt to provide broad guidance, we direct the reader to illustrations in clinical psychology \cite[]{Tacketal14}, epidemiology \cite[]{GelAur16,GelAur16a}, political science \cite[]{Tranetal18}, program evaluation \cite[]{mitchellmillennium18}, and social psychology and consumer behavior \cite[]{McSBoc17} as well as our case study in Appendix \ref{case}.

\subsection{For Editors and Reviewers}

We recommend editors and reviewers explicitly evaluate papers with regard to not only purely statistical measures but also the currently subordinate factors; this should be far superior to the status quo, namely giving consideration---often scant---to the currently subordinate factors only once some $p$-value or other statistical threshold has been reached.

One specific operationalization of this recommendation might be to incorporate consideration of these factors into various stages of the review process. For example, editors could require reviewers to provide quantitative evaluations of each factor---including domain-specific factors determined by the editor---as well as an overall quantitative evaluation of the strength of evidence as a supplement to the current open-ended, qualitative evaluations. These could then be weighted by the editors' publicly-disclosed (or even reviewers' own) importance rating of each factor. Additionally, editors could discuss and address the evaluation and importance of each factor in decision letters, thereby providing a more holistic view of the evidence.  

One might object here and call our position naive: do not editors and reviewers require some bright-line threshold to decide whether the data supporting a claim is far enough from pure noise to support publication? Do not statistical thresholds provide objective standards for what constitutes evidence, and does this not in turn provide a valuable brake on the subjectivity and personal biases of editors and reviewers? Against these, we would argue that even were such a threshold needed, it would not make sense to set it based on the $p$-value given that it seldom makes sense to calibrate evidence as a function of this statistic and given that the costs and benefits of publishing noisy results varies by field. Additionally, the $p$-value is not a purely objective standard: different model specifications and statistical tests for the same data and null hypothesis yield different $p$-values; to complicate matters further, many subjective decisions regarding data protocols and analysis procedures such as coding and exclusion are required in practice and these often strongly impact the $p$-value ultimately reported. Finally, we fail to see why such a threshold screening rule is needed: editors and reviewers already make publication decisions one at a time based on qualitative factors, and this could continue to happen if the $p$-value were demoted from its threshold screening rule to just one among many pieces of evidence.  Indeed, no single number---whether it be a $p$-value, Bayes factor, or some other statistical or non-statistical measure---is capable of eliminating subjectivity and personal biases.

Instead, we believe it is entirely acceptable to publish a paper featuring a result with, say, a $p$-value of 0.2 or a 90\% confidence interval that includes zero provided it is relevant to a theoretical or applied question of interest and the interpretation is sufficiently accurate. It should also be possible to publish a result with, say, a $p$-value of 0.001 without this being taken to imply the truth of some favored alternative hypothesis.

The $p$-value is relevant to the question of how easily a result could be explained by a particular null model, but there is no reason this should be the crucial factor in publication.  A result can be consistent with a null model but still be relevant to science or policy debates, and a result can reject a null model without offering anything of scientific interest or policy relevance.

In sum, editors and reviewers can and should feel free to accept papers and present readers with the relevant evidence. We would much rather see a paper that, for example, states that there is weak evidence for an interesting finding but that existing data remain consistent with null effects than for the publication process to screen out such findings or encourage authors to cheat to obtain statistical significance.

\subsection{Abandoning Statistical Significance Outside Scientific Publishing}

While our focus has been on statistical significance thresholds in scientific publication, similar issues arise in other areas of statistical decision making, including, for example, neuroimaging where researchers use voxelwise NHSTs to decide which results to report or take seriously; medicine where regulatory agencies such as the Food and Drug Administration use NHSTs to decide whether or not to approve new drugs; policy analysis where non-governmental and other organizations use NHSTs to determine whether interventions are beneficial or not; and business where managers use NHSTs to make binary decisions via A/B tests. In addition, thresholds arise not just around scientific publication but also within research projects, for example, when researchers use NHSTs to decide which avenues to pursue further based on preliminary findings.

While considerations around taking a more holistic view of the evidence and consequences of decisions are rather different across each of these settings and different from those in scientific publication, we nonetheless believe our proposal to demote the $p$-value from its threshold screening role and emphasize the currently subordinate factors applies in these settings. For example, in neuroimaging, the voxelwise NHST approach misses the point in that there are typically no true zeros and changes are generally happening at all brain locations at all times. Plotting images of estimates and uncertainties makes sense to us, but we see no advantage in using a threshold.

For regulatory, policy, and business decisions, cost-benefit calculations seem clearly superior to acontextual statistical thresholds. Specifically, and as noted, such thresholds implicitly express a particular tradeoff between Type I and Type II error, but in reality this tradeoff should depend on the costs, benefits, and probabilities of all outcomes.

That said, we acknowledge that thresholds---of a non-statistical variety---may sometimes be useful in these settings. For example, consider a firm contemplating sending a costly offer to customers. Suppose the firm has a customer-level model of the revenue expected in response to the offer. In this setting, it could make sense for the firm to send the offer only to customers that yield an expected profit greater than some threshold, say, zero.

Even in pure research scenarios where there is no obvious cost-benefit calculation---for example a comparison of the underlying mechanisms, as opposed to the efficacy, of two drugs used to treat some disease---we see no value in $p$-value or other statistical thresholds. Instead, we would like to see researchers simply report results: estimates, standard errors, confidence intervals, etc., with statistically inconclusive results being relevant for motivating future research.

While we see the intuitive appeal of using $p$-value or other statistical thresholds as a screening device to decide what avenues (e.g., ideas, drugs, or genes) to pursue further, this approach fundamentally does not make efficient use of data: there is in general no connection between a $p$-value---a probability based on a particular null model---and either the potential gains from pursuing a potential research lead or the predictive probability that the lead in question will ultimately be successful. Instead, to the extent that decisions do need to be made about which lines of research to pursue further, we recommend making such decisions using a model of the distribution of effect sizes and variation, thus working directly with hypotheses of interest rather than reasoning indirectly from a null model.

We would also like to see---when possible in these and other settings---more precise individual-level measurements, a greater use of within-person or longitudinal designs, and increased consideration of models that use informative priors, that feature varying treatment effects, and that are multilevel or meta-analytic in nature \cite[]{Gel15,Gel17,McSBoc17,McSBoc18}. 

\subsection{Getting From Here to There}

How do we get from here---NHST, deterministic summaries, overconfidence in results, and statistical analysis focused on reporting just some of the data---to there---statistical analysis and interpretation of results that accepts uncertainty and embraces variation and that features full reporting of results rather than focusing on whatever happens to exceed some statistical threshold?

We have offered the recommendations that we believe will serve researchers best. However, we recognize that research takes place within an institutional structure that often encourages behavior that is counter to these recommendations. Researchers respond to the expectations of funding agencies in study design and editors and reviewers in writing. Conversely, funding agencies must choose among the submissions they receive and editors can only publish papers that are submitted to them.  A careful research proposal that openly grapples with uncertainty may unfortunately lose out in the funding competition to a more traditional proposal that blithely promises 80\% power based on selected and overestimated effect sizes.  Similarly, a paper that presents all the data without making inappropriate claims of certainty may not get published in a journal that also receives submissions in which statistically significant results are presented at face value.

These institutional problems are difficult and we do not propose solutions to them. We imagine improvement will come in fits and starts, in several parallel tracks, all of which we and others have tried to contribute to in our applied and methodological research: improved statistical methods that move beyond NHST and include multilevel modeling, machine learning, statistical graphics, and other tools for analyzing and visualizing large amounts of data; applied examples using these improved methods, thereby demonstrating that it is possible to perform successful statistical analyses without aiming for deterministic results; theoretical work on the statistical effects of selection based on statistical significance and other decision criteria; and criticism of published work with gross overestimates of effect sizes or inappropriate claims of certainty.  While we recognize change will likely require institutional reform including major modifications of current practices of funding agencies and editors and reviewers, we are also optimistic that some combination of recognition of error and awareness of alternatives can also motivate change.

\section{Discussion}

In this paper, we have proposed to abandon statistical significance and offered recommendations for how this can be implemented in the scientific publication process as well as in statistical decision making more broadly. We reiterate that we have no desire to ``ban'' $p$-values or other purely statistical measures. Rather, we believe that such measures should not be thresholded and that, thresholded or not, they should not take priority over the currently subordinate factors.

While our proposal to abandon statistical significance may seem on the surface quite radical, at least one aspect of it---to treat $p$-values or other purely statistical measures continuously rather than in a thresholded manner---most certainly is not. Indeed, this was advocated by R. A. Fisher himself \cite[]{Fisher1956,GrePoo13a} as well as by other early and eminent statisticians including Karl Pearson \cite[]{StuLom09}, David Cox \cite[]{Cox77,Cox82}, and Erich Lehmann \cite[]{Leh93,Senn01}. It has also been advocated outside of statistics over the decades (see, for example, \cite{Bor19}, \cite{Eysenck60}, and \cite{SkiGueNas67}) and recently (see, for example, \cite{Drum15}, \cite{Lemetal16}, \cite{AmrKorRot17}, \cite{Gree17}, \cite{AmrGre18}). Finally, it is fully consistent with the recent American Statistical Association (ASA) Statement on Statistical Significance and $p$-values (``Principle 3: Scientific conclusions and business or policy decisions should not be based only on whether a $p$-value passes a specific threshold;'' \cite{WassLaza16}). In sum, this aspect of our proposal is part of a long literature both inside and outside of statistics over the decades that stands in direct opposition to the threshold-based status quo and the \cite{Benjetal18} proposal. 

Where our proposal might move beyond this literature is in three ways. First, we suggest that $p$-values or other purely statistical measures, thresholded or not, should not take priority over the currently subordinate factors (that said, others too have emphasized this including the recent ASA Statement which advises that ``researchers should bring many contextual factors into play to derive scientific inferences, including the design of a study, the quality of the measurements, the external evidence for the phenomenon under study, and the validity of assumptions that underlie the data analysis'' and cautions that ``no single index should substitute for scientific reasoning;'' \cite{WassLaza16}). Second, as discussed above, while we believe treating the $p$-value continuously rather than in a thresholded manner constitutes an improvement, we go further and argue that it seldom makes sense to calibrate evidence as a function of the $p$-value or other purely statistical measures. Third, we offer recommendations for authors as well as editors and reviewers for how our proposal to abandon statistical significance can be implemented in the scientific publication process as well as in statistical decision making more broadly.

Our recommendations will not themselves resolve the replication crisis, but we believe they will have the salutary effect of pushing researchers away from the pursuit of irrelevant statistical targets and toward understanding of theory, mechanism, and measurement. We also hope they will push them to move beyond the paradigm of routine ``discovery,'' and binary statements about there being ``an effect'' or ``no effect,'' to one of continuous and inevitably flawed learning that is accepting of uncertainty and variation.

\begin{appendix}

\section{Problems with Uniformly Most Powerful Bayesian Tests}\label{umpbt}

The mathematical justification underlying the \cite{Benjetal18} proposal has come under no small amount of criticism. Specifically, the uniformly most powerful Bayesian tests (UMPBTs) that underlie the proposal were introduced and defended by \cite{John13} in parallel with his call in \cite{John13b}---and now repeated in \cite{Benjetal18}---to use 0.005 as the new threshold. We see a number of concerns with UMPBTs.

First, and perhaps most relevant for the biomedical and social sciences, the UMPBT approach is deeply entrenched in the century-old Neyman-Pearson formalism of binary decisions and 0-1 loss functions. As \cite{PerPerPer13} note, ``the essence of the problem of classical testing of significance lies in its goal of minimizing Type II error (false negative) for a fixed Type I error (false positive).'' While this formalism allows for mathematical optimization under some restricted collection of distributions and testing problems, it is quite rudimentary from a decision-theoretic point of view, even to the extent of failing most purposes of conducting a sharp point null hypothesis test.

More specifically, the 0-1 loss function implicit in the NHST paradigm does not in general map, even in an approximate way, to processes of scientific learning or costs and benefits. Consequently, the logic underlying the proposal to move to a lower $p$-value threshold avoids firmly confronting the nature of the issue: any such rule implicitly expresses a particular tradeoff between Type I and Type II error, but in reality this tradeoff should depend on the costs, benefits, and probabilities of all outcomes \cite[]{GelRob14} which in turn depend on the problem at hand and which vary tremendously across studies and domains. Instead, the UMPBT is based on a minimax prior that does not correspond to any distribution of effect sizes but rather represents a worst case scenario under a set of mathematical assumptions.

Second, there is no reason for non-Bayesians to adopt UMPBTs when they can instead rely on the standard Neyman-Pearson approach to uniformly most powerful (non-Bayesian) tests. 

Third, defining the dependence of the procedure over a threshold ($\gamma$ in the notation of \cite{John13}) replicates the fundamental difficulty with the century-old Fisherian answer to hypothesis testing. To further seek a full agreement with the classical rejection region as advocated by \cite{John13} is to simply negate the appeal of a truly Bayesian approach to this issue; moreover, this agreement is impossible to achieve for realistic statistical models.

Fourth, the construction of a UMPBT relies on the assumption of a ``true'' prior, which can be criticized in a vast majority of cases and which in any case moves one away from the Bayesian paradigm: with a single and ``true'' prior, the Bayesian model becomes an errors-in-variables model.

Fifth, the argument to maximize a probability for the Bayes factor to exceed a certain threshold also moves one away from the Bayesian paradigm because: (i) it ignores the motives for running the NHST and the subsequent steps taken in decision making or inference; (ii) it further negates any prior modeling of the alternative hypothesis aimed at separating the parameter space into regions of different (prior) likelihood; (iii) it does not condition upon the actual observations but instead integrates over the observation space and hence may fall afoul of the likelihood principle; (iv) it posits a single and fixed threshold $\gamma$ for rejecting the null when there is no reason for $\gamma$ not to depend on the observed data, as also argued above; (v) the maximization step eliminates the role of the prior distribution, as also argued above; (vi) in the rare one-dimensional settings where the maximization step can be conducted in closed form, the solution is a distribution with finite support; (vii) in the event the null hypothesis is rejected, the uniformly most powerful prior (or alt-prior) corresponding to the alternative cannot be used as such in subsequent inference but must instead be replaced with a regular prior over the whole parameter space---a strong violation of Bayesian coherence.

Sixth, speaking more generally, the concept of uniformly most powerful priors (and tests) does not easily extend to multivariate settings and even less to realistic cases that involve complex null hypotheses that contain nuisance parameters. The first solution proposed in \cite{John13}, to integrate out the nuisance parameters in the null hypothesis using a specific prior distribution, falls short of solving the issue of ``objective Bayesian tests.'' The second solution, namely to replace the unknown nuisance parameters with standard estimates, stands even farther from a Bayesian perspective. 

Indeed, the Bayes factor itself is a consequence of the rudimentary Neyman-Pearson formalism, which as such caters to the issue of statistical significance. A discussion of the difficulties with this from a Bayesian perspective is provided in \cite{Kamaetal14}, with a proposal of setting the hypothesis problem as one of mixture estimation.

Seventh, \cite{John13} contains very little support for the asymptotic relevance of the approach, beyond the limiting normal distribution of the uniformly most powerful log Bayes factor and the convergence of the support to the ``true'' value of the parameter.

In closing, we note that many of our criticisms of the \cite{John13} approach relate to the fact that it falls short of being truly Bayesian. However, we do not mean to say that hypothesis testing must be done in a Bayesian manner. Rather, we emphasize this because, to the extent that the \cite{John13} approach loses its Bayesian connection, it also loses a Bayesian justification for the 0.005 rule. Consequently, 0.005 becomes just another arbitrary threshold, justified by some implicit tradeoff between false positives and negatives which we think does not make sense in any absolute and acontextual way.

\section{Case Study}\label{case}

In the context of a hypothetical case study on the effects of sodium on blood pressure, we discuss how authors as well as editors and reviewers might follow our recommendation to demote the $p$-value from its threshold screening role and instead treat it continuously along with the currently subordinate factors---related prior evidence, plausibility of mechanism, study design and data quality, real world costs and benefits, novelty of finding, and other factors that vary by research domain---as just one among many pieces of evidence.

We recommend authors use the currently subordinate factors to motivate their data collection, statistical analysis, interpretation of results, writing, and related matters. In this example, the authors might consider {\em related prior evidence} that indicates the importance of blood pressure as a marker for healthy arteries, suggests the role of sodium in hemodynamics, and so forth. This evidence might also reveal a {\em plausible mechanism}, namely to excrete excess sodium the body must increase blood pressure. 

In terms of {\em study design and data quality}, the authors might consider various possibilities for data collection. How should they recruit subjects? Should they randomize them to a low-sodium versus high-sodium diet? Or should they track them longitudinally, say via routine annual checkups over the course of years? Or is such data already available from some prior study? When and how often should sodium and blood pressure be measured? And how? The authors might measure sodium through a dietary recall questionnaire (noisy), through asking participants to maintain a food diary (somewhat less noisy), or through collection of urine to measure urinary sodium excretion (precise but restricted to a limited time point). Likewise, for blood pressure, they might rely on measurements conducted by someone convenient like friends or family members of the subjects who likely do not possess formal clinical training (noisy) or by paid clinicians instructed on the proper protocol for blood pressure measurement (precise but expensive).

Suppose the authors hypothesize a positive association between sodium consumption and high blood pressure. For the moment, let us assume that---while eschewing the NHST paradigm and the $p$-value thresholds intrinsic to it---the authors nonetheless perform a statistical analysis that results in a $p$-value. Further, let us assume they obtain a $p$-value of 0.001. How should this impact their interpretation of results, writing, and statistical decision making more broadly? Certainly, they have gained support for their hypothesis. However, can they conclude sodium is associated with---or even causes---high blood pressure as they would under the NHST paradigm?

Well, it would depend on the context and limitations of the {\em study design and data quality}. For example, supposing the study took place in Japan, perhaps the association exists in the Japanese subject population studied but does not in European populations whether because of some genetic differences between the two populations or because of some dietary differences (e.g., dietary sodium levels are much higher among Japanese so the association might not hold in levels typical among Europeans).

In terms of a causal interpretation, this would depend on {\em related prior evidence}, {\em plausibility of mechanism}, and {\em study design and data quality}. If prior studies show consistent and strong associations between sodium consumption and blood pressure, if evidence from physiological studies and animal models are consistent with an effect of sodium consumption on blood pressure, or if sodium levels were randomized, this increases the support for a causal role of sodium in increasing blood pressure.

Given, say, that the causal interpretation holds and holds broadly, the authors could then consider clinical significance, that is, {\em real world costs and benefits}. This depends not at all on a $p$-value but on the estimates of the magnitude of the effect---not only on blood pressure but also on downstream outcomes such as cardiovascular disease---as well as the uncertainty in them. It also depends on the costs of potential interventions such as lower sodium diets and drugs. They could also discuss {\em novelty of finding} in light of all of the above.

Now, let us assume they had instead obtained a $p$-value of 0.2. Can they conclude sodium is not associated with high blood pressure as they would under the NHST paradigm? Again, this would depend on all the factors discussed above. For example, perhaps the association does not exist in the Japanese population but does in European ones and so on.

There are two key points in this. First, results need not first have a $p$-value or some other purely statistical measure that attains some threshold before consideration is given to the currently subordinate factors. Instead, and as illustrated above, statistical measures should be considered along with the currently subordinate factors as just one among many pieces of evidence and should not take priority thereby yielding a more holistic view of the evidence. Second, statistical measures should be treated continuously in this more holistic view of the evidence. Specifically, a lower $p$-value constitutes continuously stronger evidence---and this holds regardless of the level of the $p$-value. Further, this continuously stronger evidence can be balanced along with the strengths and weaknesses of the currently subordinate factors in assessing the level of support for a hypothesis.

Of course, we believe not only that the authors' statistical analysis should not be restricted to the NHST paradigm and the $p$-value thresholds intrinsic to it but also that it need not---and often should not---even result in a $p$-value (i.e., because it seldom makes sense to calibrate evidence as a function of the $p$-value). As noted, we recommend authors report all of their data and relevant results rather than focusing on single comparisons that attain some $p$-value or other statistical threshold. In this context, this might involve modeling the association between sodium and blood pressure as a function of additional health and dietary variables, demographic variables, and geography using a multilevel model. Such a model would not yield one single $p$-value thereby encouraging dichotomous declarations of truth or falsity---binary statements about there being ``an effect'' or ``no effect.'' Instead, it would yield many estimates that vary based on, for example, health and dietary variables, demographic variables, and geography, as well as the uncertainty in these estimates. Indeed, by accepting uncertainty and embracing variation in effects, the authors would uncover and present a much richer and more nuanced story about the association between sodium and blood pressure.

Turning to editors and reviewers, we recommend they explicitly evaluate papers with regard to not only purely statistical measures but also the currently subordinate factors. How might this work? We envision it would be rather similar to the above but in reverse. Specifically, editors and reviewers evaluating the authors' paper on sodium and blood pressure would systemically assess, and possibly even indicate the weight they assign to, each of the following: How does the paper fit in with and build upon {\em related prior evidence}? Is the {\em mechanism plausible}? Are the {\em study design and data quality} sufficient to justify the conclusions? What are the implications in terms of {\em real world costs and benefits}? How {\em novel} are the findings? And, of course, how appropriate are the statistical analyses employed and how strong is the statistical support, whether in the form of a $p$-value or some other measure, resulting from these analyses?

In this more holistic view of the evidence, statistical measures are just one among many pieces of evidence considered by editors and reviewers and do not take priority. Of course, this does not mean that they cannot or will not strongly impact or alter their evaluation decisions. For example, in the context of the authors' paper on sodium and blood pressure, strong statistical support, whether in the form of a low $p$-value or otherwise, for a finding that sodium consumption is associated with low blood pressure---the direction opposite of that indicated by prior evidence---in the context of a high quality study design featuring large samples and precise measurements might be deemed more novel and worthy of publication than if the statistical support had been weaker or if the finding was in the same direction as that indicated by prior evidence. 

In sum, authors as well as reviewers and editors need not use statistical significance as a lexicographic decision rule.  Results need not first have a $p$-value or some other purely statistical measure that attains some threshold before consideration is given to the currently subordinate factors. Instead, treated continuously, statistical measures should be considered along with the currently subordinate factors as just one among many pieces of evidence and should not take priority thereby yielding a more holistic view of the evidence.

\end{appendix}

\bibliographystyle{natbib}
\bibliography{abref}

\end{document}